\documentclass[]{JHEP3} % 10pt is ignored!

\JHEPspecialurl{http://www.pnas.org/cgi/content/full/102/26/9150}

\usepackage{epsfig,multicol,bbm}

\newcommand\fverb{\setbox\pippobox=\hbox\bgroup\verb}
\newcommand\fverbdo{\egroup\medskip\noindent%
            \fbox{\unhbox\pippobox}\ }
\newcommand\fverbit{\egroup\item[\fbox{\unhbox\pippobox}]}
\newbox\pippobox
\title{The Bacterial Chemotactic Response Reflects a Compromise Between
Transient and Steady State Behavior}

\author{Damon A. Clark
        and Lars C. Grant\\
    Department of Physics, Harvard University\\ 17 Oxford Street, Cambridge MA 02138, USA\\
    E-mail: \email{daclark@fas.harvard.edu}, \email{lgrant@fas.harvard.edu}}

\preprint{\hepth{9912999}}  % OR: \preprint{Aaaa/Mm/Yy\\Aaa-aa/Nnnnnn}
\abstract{Swimming bacteria detect chemical gradients by performing
temporal comparisons of recent measurements of chemical
concentration. These comparisons are described quantitatively by the
chemotactic response function, which we expect to optimize
chemotactic behavioral performance. We identify two independent
chemotactic performance criteria: in the short run, a favorable
response function should move bacteria up chemoattractant gradients,
while in the long run, bacteria should aggregate at peaks of
chemoattractant concentration. Surprisingly, these two criteria
conflict, so that when one performance criterion is most favorable,
the other is unfavorable. Since both types of behavior are
biologically relevant, we include both behaviors in a composite
optimization that yields a response function that closely resembles
experimental measurements. Our work suggests that the bacterial
chemotactic response function can be derived from simple behavioral
considerations, and sheds light on how the response function
contributes to chemotactic performance.}

\keywords{chemotaxis, optimization, strategy}

\dedicated{This article has been published: \textit{PNAS}
\textbf{102}(26): 9150-9155. The published version is available at:
http://www.pnas.org/cgi/content/full/102/26/9150.}

\begin{document}

%\maketitle  IS IGNORED %%%%%%%%%%%

\section{Introduction}

The bacterium \textit{E. coli} moves up gradients to regions of high
chemoattractant concentration by performing a biased random walk.
The random walk consists of alternating runs (periods of forward
movement) and tumbles (sudden reorientations) that arise from
changes in flagellar rotation \cite{brown,larsen}. When the flagella
rotate counter clockwise, they form a bundle and the bacterium swims
more or less in a straight line at a roughly uniform speed. When one
or more flagella rotate clockwise, they leave the bundle and the
bacterium tumbles, randomly re-orienting itself \cite{mcnab,turner}.
Bacteria bias the random walk by modulating the run duration in
response to measurements of chemoattractant concentration that are
made at the cell surface \cite{stock, bray}. They do not perform
spatial comparisons between points along the cell body because of
the fast diffusion across such short distances \cite{bergpurcell}.

The chemotactic response function describes how bacteria process
concentration measurements to produce their behavioral run-biasing
decisions. It has been measured experimentally by monitoring the
rotation of single flagella on bacteria stimulated by instantaneous
chemoattractant pulses \cite{segall}. The empirical response
function is biphasic: the pulse provokes an immediate brief
elevation of the counter clockwise probability immediately followed
by a longer depression. We expect that the shape of the chemotactic
response function should deliver optimal behavioral performance.

We consider the chemotactic behavior of a bacterium at some specific
position on a gradient of attractant.  As it wanders up and down the
gradient, the distribution of its positions approaches a steady
state. We choose performance criteria that quantitatively
characterize the performance of the bacterium at early times in the
non-steady state regime and at late times in steady state. Both of
these regimes are biologically relevant.  If the system navigated by
the bacterium is small compared to the distance the bacterium could
explore in the time between cell divisions (an example is bacterial
aggregation into clusters \cite{oudenaarden}), then it is the steady
state behavior that matters most to the bacterium. If, however, the
system is large -- more than a few millimeters in size -- or varies
in time, the bacterium will not come to a steady state before
dividing, and a single cell might never reach a steady state.
Bacteria have no \textit{a priori} knowledge of the size of their
system, so their chemotactic strategy should benefit them in either
the steady state or non-steady state regime.  Following foraging
theory \cite{parker, foraging_book}, we will assume that the
chemotactic strategy maximizes the attractant seen by the bacterium
on the timescale of bacterial divisions.

Our first performance criterion reflects the expected velocity of
bacteria at early times, before they have reached the boundaries of
the system. It is quantified by $\mathcal{T}$, a measure of the
early time transient velocity of bacteria with a given response
function.  This was previously calculated by de Gennes
\cite{degennes}. Optimizing $\mathcal{T}$ leads to a single-lobed
response function, which causes a population of bacteria to have a
transient average velocity up gradients at early times. Contrary to
intuition, this optimization leads to an unfavorable steady state
distribution with bacteria accumulated in regions of low attractant.

The second performance criterion, $\mathcal{S}$, quantifies how
strongly the bacteria aggregate about chemoattractant maxima when in
steady state. Optimizing $\mathcal{S}$ leads to a bacterium that has
a mean velocity down gradients at early times, but whose position
distribution peaks at high concentrations at long times. The two
performance criteria conflict: when one is maximal, the other is
unfavorable. If both performance criteria are used to calculate the
response function, the theoretical function closely matches the
empirical biphasic bias curve measured by Segall, Block, and Berg
\cite{segall}. The optimization procedure explains the curve's
structure.

Our work contributes to a body of theoretical investigations of
bacterial chemotaxis. Schnitzer \textit{et al.} \cite{schnitzer}
used Monte Carlo simulations to confirm the favorable performance of
a biphasic response function compared to a monophasic one.  Our
approach supports their end result, though we show that aggregation
can occur without a positive lobe on the response function.
Schnitzer \cite{schnitzer_PRE} also adopted a kinetic approach and
derived results about steady state behavior in a variety of cases.
He distinguished between `nonadaptive pseudochemotaxis' and `true
adaptive chemotaxis'. In contrast, our approach emphasizes both
transient and steady state behavior in evaluating chemotaxis. In a
different approach, Strong \textit{et al.} \cite{strong} adopted a
deterministic model for tumbling and examined optimality in the
presence of signal noise. Work by de Gennes \cite{degennes} focused
on the mean bacterial velocity due to a given response function. We
show that this mean velocity only applies at early times, and we
extend the framework used by de Gennes to examine steady state
performance and performance optimization.

%----------------------------------------------------------------------------
\section{Model Details}\label{model}
%\textbf{Model Details}

We adopt the stochastic framework used by de Gennes \cite{degennes}.
In this model, bacteria continuously modulate their instantaneous
probability of tumbling as a function of a differential weighting of
past measurements of chemoattractant concentration. The differential
weighting constitutes the chemotactic response function, $R(t)$.

We assume that the chemical landscape is static and that
chemoattractant concentration is defined at every point by a
function $c(x)$. Bacteria swim along individual paths $x(t)$ at a
uniform speed $v$. The probability, $P$, that a bacterium tumbles in
an interval between $t$ and $t+dt$ is dictated by its entire
previous path, the chemical landscape, and the chemotactic response
function:
\begin{equation}
\label{basicP} P[x(t');t]dt = \frac{dt}{\tau} \left[ 1 -
\int_{-\infty}^t dt'' R(t-t'')c(x(t'')) \right]
\end{equation}
where $\tau$ is the mean run duration in the absence of a
perturbation. In a uniform concentration, this model describes
tumbling as an unbiased Poisson process with a constant rate of
tumbling $1/\tilde{\tau}$ given by $(1-c\int R(t)dt)/\tau$. In a
concentration gradient, $P$ depends on the bacterium's history. By
choosing particular forms of the response function, bacteria can
bias their random walk so that they climb gradients and remain in
regions of high $c$.

We will consider first order perturbations of the Poisson process by
defining $R(t) \propto \alpha/\tau$, with $\alpha$ small such that
$\int_{-\infty}^t dt' R(t-t')c(t') \ll 1$.  The constant $\alpha$
has units of inverse concentration. We expand equations as power
series of such integrals and discard higher order terms involving
products of such integrals.  Equation (\ref{basicP}) can itself be
regarded as the first order expansion of some monotonic function of
$\int_{-\infty}^t dt'' R(t-t'')c(x(t''))$ that remains positive for
all concentrations.

In our analysis, we neglect the effects of noise due to fluctuations
of $c(x)$.  Noise averages to zero in all our first order
expansions.  The first noise contribution that does not average to
zero is proportional to the variance of the concentration, and is of
order $\alpha^2c/V$, where $V$ is the cell volume.  To neglect this
term with respect to the first order term, we require that $\alpha
\ll V$. The experimental conditions described by Segall
\cite{segall} correspond to the regime in which bacterial responses
are linear and the bacteria can detect $c$ without being overwhelmed
by noise.

Berg and Purcell \cite{bergpurcell} argued that measurement
integration times of about 1 second account for observed sensitivity
to concentrations and gradients in the presence of noise.  The
response functions resulting from our analysis vary on the timescale
$\tau$, about 1 second, so they will display biological sensitivity
without explicitly requiring long integration times.

We assume that the length scale of variations in the concentration
gradient is longer than the average run length, so that over one run
the gradient appears linear. We consider bacteria in one dimension
and assume that tumbles are instantaneous and randomize orientation.
Of course, real bacteria navigate in three dimensions and their run
directions are not completely decorrelated by tumbles
\cite{brown,turner}. Further, in real bacteria, runs directed up
attractant gradients lengthen, while those directed downwards are
the same length as runs in constant concentrations \cite{brown}.
Nonetheless, this simplified model gives insight into real bacterial
behavior.

%----------------------------------------------------------------------
\section{Transient Chemotaxis}
\label{runtimes}
%\textbf{Moving up a Gradient}

The strategic goal of a bacterium navigating a chemoattractant
landscape is arguably as simple as producing an average velocity up
the attractant gradient. De Gennes showed that a mean velocity can
be produced when, after a tumble, a run up the gradient lasts longer
than a run down the gradient \cite{degennes}. For a population of
bacteria starting at the same position, the expected velocity at
early times will be:
\begin{equation} \label{meanv_def}
\overline{v}\simeq v \frac{\overline{\Delta t}}{2\tau}
\end{equation}
where $\Delta t$ is the difference in run times moving up and down
the gradient and the bars are averages over possible trajectories.

The model presented in Equation (\ref{basicP}) dictates that the
probability of next tumbling at time $t_f$ after having previously
tumbled at time $t_0$ is
\begin{equation} \label{eqPdef}
P(t_f| t_0) = P[x(t'');t_f] \exp \left\{-\int_{t_0}^{t_f}
P[x(t'');t']dt' \right\}
\end{equation}
where $P[x(t'');t_f]$ is the probability of tumbling at time $t_f$,
given a path $x(t'')$.

Following De Gennes \cite{degennes}, we consider the expected time
until the next tumble:
\begin{equation}\label{eqforwardint}
\overline{t}(x_0) = \int_{t_0}^\infty dt_f (t_f-t_0) P(t_f| t_0)
\end{equation}
where the average is taken over possible future trajectories and the
bacterium is at $x_0$ at time $t_0$. We define
$\overline{t}^\pm(x_0)$ as the mean time until the next tumble for
bacteria moving up ($+$) and down ($-$) the gradient. We expand
$c(x(t))$ into $c(x_0) \pm v(\nabla c)(t-t_0)$, where $v$ is the
constant speed of a run. After expanding in $\alpha$, and then using
the identity $R(t)= \int_0^{\infty} R(s) \delta (s-t) dt$ where
$\delta (s-t)$ is a Dirac delta function, we find de Gennes's result
that:
\begin{eqnarray}
\label{eqtimediff}  \overline{v} = \frac{v}{2\tau}
(\overline{t}^+(x_0) - \overline{t}^-(x_0)) =  v^2 \tau \nabla
c(x_0) \int_0^{\infty} e^{-t/\tau}R(t)dt
\end{eqnarray}
Figure \ref{figintill} illustrates this integral over future paths.
For bacteria to behave most favorably at early times, $\overline{v}$
should be maximal.

\FIGURE{\epsfig{file=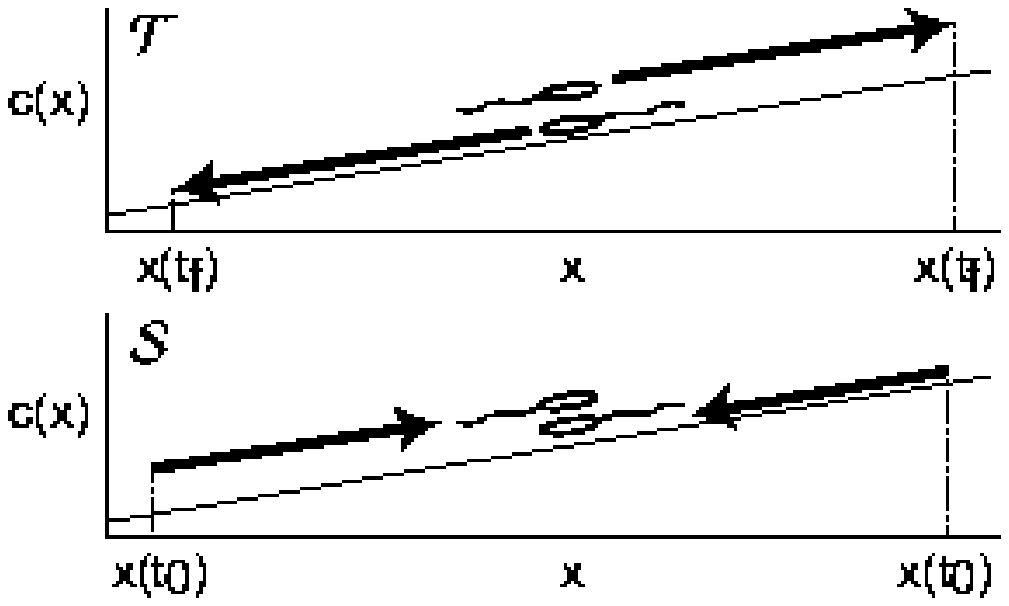,width=9cm}
        \caption{The top figure illustrates the
integration in the expression for $\mathcal{T}$ (Equation
(\ref{eq_T_definition})).  Two bacteria that have both just tumbled
are considered as they move in different directions along the
gradient until they tumble again at position $x(t_f)$. The bottom
figure illustrates the integration in the expression for
$\mathcal{S}$ (Equation (\ref{S_definition})). In this case, two
bacteria meet that last tumbled at points $x(t_0)$. One finds the
expectation of their respective tumbling probabilities,
$\overline{P}^\pm$, by averaging over possible histories.}
    \label{figintill}}

To gain intuition about this mean velocity, consider 1000 bacteria
all taking exactly average steps, beginning at a point $x_0$ on an
infinitely long gradient, as illustrated in Figure \ref{figpaths}.
Initially, 500 move up the gradient until the time $\overline{t}^+$,
while 500 move down until time $\overline{t}^-$. The average
position of the bacteria is simply $x_0$ until time
$\overline{t}^-$, when the 500 moving down split into 250 moving up
and 250 moving down. From this time until the up-moving bacteria
tumble at time $\overline{t}^+$, the mean position of the bacteria
moves up at $v/2$. This phenomenon is repeated after every tumble,
creating the mean velocity up the gradient (Equation
(\ref{meanv_def})). In Figure \ref{figpaths}, this mean velocity is
reflected by the thick tail of bacteria on the up-moving branch and
the thinner tail on the down-moving branch. When the upward moving
tail encounters a boundary on the system, bacteria are forced to
tumble and the mean velocity up the gradient dies away as the
bacteria move toward their steady state distribution. Figure
\ref{figsim}\textsf{\textbf{b}} shows the results of a simulation
that demonstrates this transient behavior.

We divide out the constants in (\ref{eqtimediff}) and introduce the
dimensionless performance measure
\begin{equation} \label{eq_T_definition}
\mathcal{T}[R(t)]=\frac{\overline{t}^+ - \overline{t}^-}{2 \alpha v
\tau^2 \nabla c}=\frac{1}{\alpha}\int_0^{\infty} e^{-t/\tau}R(t)dt
\end{equation}
to quantify the transient chemotactic behavior at early times.  This
quantity is an overlap integral of $R(t)$ against a performance
kernel $K_{\mathcal{T}}(t)={1\over\alpha} e^{-t/\tau}$. The
performance kernel shows the effect of the response function on the
mean velocity at early times. The form of this kernel can be
understood qualitatively.  The mean velocity is proportional to the
difference in run times between two bacteria with the same starting
point, moving in different directions (see Figure \ref{figintill}).
As up- and down-moving bacteria move away from each other, the
difference in the concentrations they measure grows until the
bacteria tumble. Therefore a response that weights $c(t)$ heavily in
the immediate past will contribute more to increasing $\mathcal{T}$
than a weighting further in the past where concentration differences
were smaller. This is why the performance kernel prefers recent
weighting. The shape of the performance kernel matches simulations
of the model system (Figure \ref{figsim}\textsf{\textbf{c}}).  The
exponential decrease in influence of $R(t)$ on $\mathcal{T}$ is due
to the exponential run length of the unperturbed Poisson process.
Note that this heuristic argument is not strongly dependent on the
form of $P$ chosen in Equation (\ref{basicP}). Any positive
decreasing function of $\int_{-\infty}^t dt'' R(t-t'')c(x(t''))$
would yield a kernel that weights the most recent measurements most
heavily.

\FIGURE{\epsfig{file=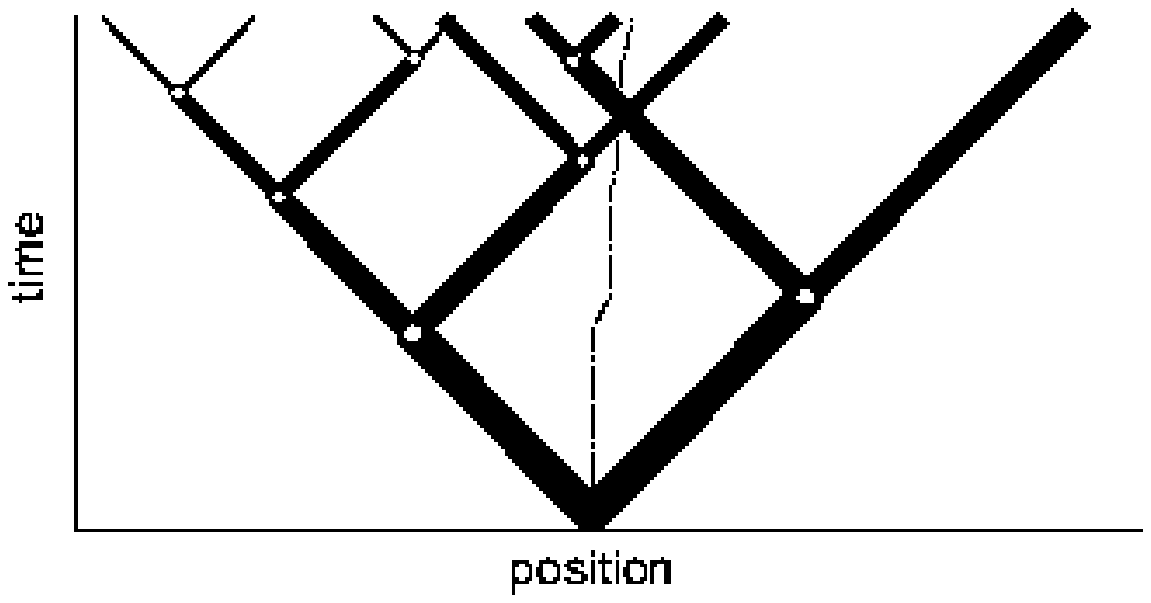,width=9cm}
        \caption{Solid lines indicate possible paths taken
by bacteria that all execute exactly average paths; line thickness
gives a sense of the probability weighting of each path segment. The
chemoattractant gradient in this case is positive and
$\overline{t}^+ > \overline{t}^-$. The dotted line shows the average
position over time: it moves to the right, indicating an expected
velocity up the gradient.  Note that after the time elapsed in this
figure, more bacteria on average will have reached the furthest
right point than the furthest left point, just because they have
tumbled less frequently.}
    \label{figpaths}}

We can maximize $\mathcal{T}$ over a constrained set of response
functions. We assume the response function to be finite and to decay
to $0$ at large $t$. The simplest way to include both restrictions
is to hold the integral $\int_0^\infty R^2(t)dt$ constant. This
amounts to maximizing over a set of response functions that have the
same root-mean-squared deviation from $0$.  We impose the constraint
\begin{equation}
\int_0^{\infty}R^2(t)dt=\alpha^2/\tau
\end{equation}
and maximize $\mathcal{T}$ subject to this constraint by using a
Lagrange multiplier and taking a functional derivative:
\begin{equation} \label{funcderiv}
\frac{\delta}{\delta R(t)} \bigg[\mathcal{T} + \lambda \bigg(
\frac{\tau}{\alpha^2}\int_0^\infty R(t')^2 dt' - 1 \bigg)\bigg] = 0
\end{equation}

Solving this condition, we calculate the optimized response function
\begin{equation} \label{eqRmaxt}
R_\mathcal{T}(t)=\frac{\alpha}{\tau}N_\mathcal{T}\exp \{-t/\tau\}
\end{equation}
where $N_\mathcal{T}$ is a normalization constant. This response
function is proportional to the performance kernel
$K_{\mathcal{T}}(t)$ used to determine $\mathcal{T}$; it is positive
everywhere but weighted towards most recent times (shown in Figure
\ref{figdata}\textsf{\textbf{a}}). Using this response function,
bacteria moving up and down the gradient are progressively less and
more likely to tumble, respectively. Given a particular tumbling
position $x$, this results in maximally longer runs up the gradient
than down it.  A similar effect has been termed `pseudochemotaxis'
\cite{lapidus}.  We call it `transient chemotaxis' because unlike in
pseudochemotaxis, $P[x(t');t]$ in transient chemotaxis has a history
dependence, and moreover we argue that short-term performance is
relevant for bacteria in large chemical gradients.

Surprisingly, although $R_\mathcal{T}(t)$ maximizes the expected
velocity up the gradient, it leads to an unfavorable steady state
distribution in which the bacteria spend more time in low
chemoattractant concentration regions. The simulation in Figure
\ref{figsim}\textsf{\textbf{b}} shows the initial favorable
transient velocity and the unfavorable steady state for an
all-positive response function. This counter-intuitive result can be
explained as follows. Imagine two bacteria passing each other on a
linear concentration gradient (see Figure \ref{figintill}). The one
heading down the gradient has high $c$ in its past, so its value of
$\int R_\mathcal{T}(t-t')c(x(t'))dt'$ is larger on average than that
of an upward-moving one at the same position. Equation
(\ref{basicP}) then shows that the probability of tumbling is
\textit{lower} for the bacterium moving down the gradient.  Since
this is true at all points on the gradient, more bacteria will
accumulate in the low concentration areas. The unfavorable steady
state of a positive response function was previously shown in
numerical simulations \cite{schnitzer} and noted in Schnitzer's
analysis \cite{schnitzer_PRE}.

Initial velocity need not indicate the eventual steady state
distribution, as the following thought experiment shows (discussed
by Lapidus \cite{lapidus}, Schnitzer \textit{et al.}
\cite{schnitzer}, and Schnitzer \cite{schnitzer_PRE}). Consider a
closed tube containing a gradient in density of steel wool. At one
end of the tube, mean free paths of a molecule are short, while at
the other end they are long. After each collision, because of the
gradient in the wool, a molecule has an expected net displacement
towards the sparse end of the tube. In steady state, however, gas
molecules are distributed evenly throughout the free volume of the
tube. Therefore, although the expected net displacement after each
collision creates an initial mean velocity towards the sparse end,
it does not determine the steady state distribution. For gas
molecules, the collision probability is determined by a particle's
instantaneous position. For bacteria using an all-positive response
function, both $\overline{t}^+$ and $\overline{t}^-$ are longer in
higher $c$ regions because $\int_0^\infty R_\mathcal{T}(t)dt \not=
0$, making path length depend on position. It is the
history-dependence of $R_\mathcal{T}$ that causes the bacteria to
aggregate in regions of low $c$.

%-------------------------------------------------------------------------
\section{Steady State Bacterial Distribution}
% \section{General Steady State Distribution}
\label{steadystate}
%\textbf{General Steady State Distribution}

Here we show how the steady state distribution of bacteria depends
on expected tumbling rates. The expected tumbling rate for a
bacterium at position $x$ depends on whether it is moving up or down
the gradient, and is given by $\overline{P}^+(x)$ or
$\overline{P}^-(x)$, respectively, where bars are averages over
possible histories ending at $x$. In steady state these averages
will not be functions of $t$.

In steady state, bacterial flux is zero and the bacterial steady
state concentration, $b(x)$, can be written in terms of the
probabilities $\overline{P}^\pm(x)$ (see the Supplementary
Information):
\begin{equation}\label{eqssfundamental} b(x)=b_0 \exp\left\{\int_0^x \frac{dx'}{2v}
\left( \overline{P}^-(x') - \overline{P}^+(x')\right) \right\}
\end{equation}
This reproduces a more general result derived in
\cite{schnitzer_PRE}.

With net flux zero, the number of upward-moving bacteria must equal
the number of downward-moving bacteria at any point $x$. If
$\overline{P}^+(x) \not= \overline{P}^-(x)$, then the
\textit{fraction} of bacteria passing through a point from the left
will not equal that passing through from the right.  In order to
keep the \textit{number} fluxes equal, the number of bacteria on
each side of that point must be different. Maintaining this balance
generates the form of the distribution in Equation
(\ref{eqssfundamental}). When the tumbling rate is higher for
down-moving bacteria arriving at point $x$, bacteria aggregate at
the top of the gradient in steady state.

\clearpage
\FIGURE{\epsfig{file=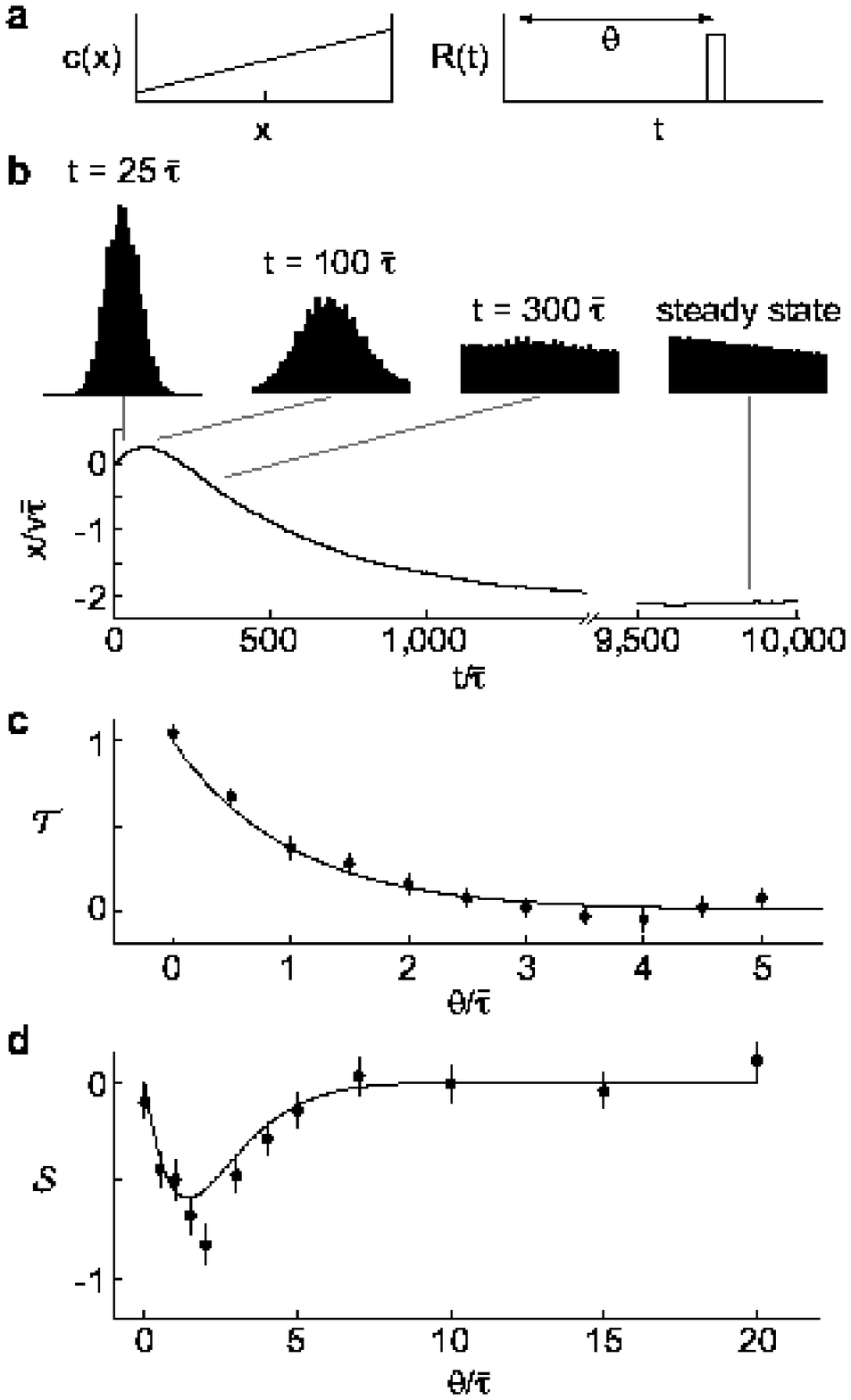,width=9cm}
        \caption{We performed discrete time simulations of
the model on a positive concentration gradient with reflective
boundary conditions to see the result of different $R(t)$ on
transient and steady state behaviors.  Bacteria were released from
the center of the gradient (\textsf{\textbf{a}}, left) and evolved
until they arrived at a steady state distribution.  $R(t)$ was
chosen to weight positively only at $\theta$ seconds before the
current time, $t$ (\textsf{\textbf{a}}, right); that is, it weights
only $c(t-\theta)$. It was further chosen so that the maximum
perturbation from the average tumbling probability was 30\%.
\textsf{\textbf{b}} In a gradient of length $60 v\tilde{\tau}$,
bacterial distributions and the mean position of bacteria were found
using a response function with $\theta=\tilde{\tau}$, where
$\tilde{\tau}$ is the run duration averaged over the box.  At early
times, bacteria are clustered and have a mean velocity up the
gradient. After the bacteria hit the boundary, they approach a
steady state peaked at low $c$.  Note that more bacteria have
reached the righthand wall than left hand wall at
$t=100\tilde{\tau}$.  For this response function, $\mathcal{T}>0$
and $\mathcal{S}<0$; both results are reflected in the bacterial
behavior. \textsf{\textbf{c}} We varied $\theta$ and calculated
$\mathcal{T}$ from the initial slope of the lower plot in
\textsf{\textbf{b}}.  The result shows the contribution of
$R(\theta)$ to $\mathcal{T}$.  The solid line is the transient
performance kernel, $K_\mathcal{T}$, derived in the text.
\textsf{\textbf{d}} In a short length scale gradient
(4$v\tilde{\tau}$), we varied $\theta$ and calculated $\mathcal{S}$
from the bacterial distributions at long times. This shows the
contribution of $R(\theta)$ to $\mathcal{S}$. The solid line is the
steady state performance kernel, $K_\mathcal{S}$, derived for a
similar situation (see Supplementary Information). Error bars in
\textsf{\textbf{c}} and \textsf{\textbf{d}} are 1 SEM.}
    \label{figsim}}
\clearpage

%---------------------------------------------------------------------------
%\section{Bacterial Distributions at Long Times}
%\label{steadystate}
%\textbf{Bacterial Distributions at Long Times}

We now express the tumbling probabilities in terms of $R(t)$. In
order to calculate $\overline{P}^\pm(x)$ we must consider all
possible histories of bacteria reaching point $x$.  Histories and
instantaneous tumbling probabilities both depend on $R(t)$, so the
difference $\overline{P}^-(x) - \overline{P}^+(x)$ that governs
steady state aggregation can be expressed in terms of the response
function. By integrating over paths for bacteria arriving at $x$
(details of the derivation are in the Supplementary Information), we
find that
\begin{equation} \label{eqPtdiff}
\overline{P}^-(x) - \overline{P}^+(x) =2v \nabla c(x)
\int_0^{\infty} -(t/\tau + t^2/2 \tau^2)e^{-t/\tau} R(t) dt
\end{equation}

This integral should be positive to obtain an advantageous steady
state distribution with more bacteria at high concentrations. The
$x$ dependence in Equation (\ref{eqPtdiff}) comes through the
$\nabla c$ factor, which integrates immediately to $c(x)$, giving
the steady state distribution:
\begin{equation}
b(x)=b_0 \exp \bigg\{ c(x) \bigg( \frac{
\overline{P}^-(x)-\overline{P}^+(x) }{2 v\nabla c(x)}
  \bigg) \bigg\}
\end{equation}
The quantity in round brackets does not depend on $x$. We introduce
the dimensionless version of this quantity,
\begin{equation} \label{S_definition}
\mathcal{S}[R(t)] = \frac{\overline{P}^-(x)-\overline{P}^+(x)}{2 v
\alpha \nabla c(x)} = \frac{1}{\alpha}\int_0^{\infty} -(t/\tau +
t^2/2 \tau^2)e^{-t/\tau} R(t) dt
\end{equation}
as a performance measure of the steady state distribution.  This is
an overlap integral, with a performance kernel
$K_{\mathcal{S}}(t)=-{1\over\alpha}(t/\tau + t^2/2
\tau^2)e^{-t/\tau}$. A response with large $\mathcal{S}$ yields a
steady state distribution with the bacteria aggregated favorably in
high $c$ regions.

When $\mathcal{S}$ is maximized by the same procedure used in
Equation (\ref{funcderiv}), one finds a response function
\begin{equation}\label{eq_R_S_def}
R_\mathcal{S}(t)=-
\frac{\alpha}{\tau}N_\mathcal{S}(t/\tau+t^2/2\tau^2) \exp\{-t/\tau\}
\end{equation}
which is negative everywhere, zero at $t=0$ and at long times, and
peaked at $t=\tau\sqrt{2}$ (see Figure
\ref{figdata}\textsf{\textbf{a}}). The negative values of this
response function mean that bacteria moving down the gradient at
point $x$, with high concentrations in their past, have higher
tumbling probabilities than bacteria moving up the gradient at $x$,
with lower concentrations in their past.

Because $R_\mathcal{S}(t)$ is negative, it results in
$\overline{t}^-(x) > \overline{t}^+(x)$, creating a transient
velocity down the gradient at early times. Although this response
function gives a beneficial steady state distribution, it yields
detrimental behavior at early times.

One can understand the steady state performance kernel
qualitatively. The performance measure $\mathcal{S}$ considers the
difference in tumbling probability between two bacteria at the same
point in space but coming from opposite directions (see Figure
\ref{figintill}). In this case, measurements of $c$ are most
different in the past, while the most recent concentration
measurement, $c(x)$, is the same for both bacteria. This weighting
is reflected in the performance kernel $K_{\mathcal{S}}$ and in the
optimal response $R_\mathcal{S}(t)$, in which concentrations in the
past are more heavily weighted. Concentration measurements in the
more distant past could have been made where $\nabla c$ was
different from the current $\nabla c$ and cannot be reliably related
to the current gradient.  Therefore, such distant information is not
useful for making run-biasing decisions and is not weighted heavily
by the kernel \cite{macnab_koshland}. Figure
\ref{figsim}\textsf{\textbf{d}} shows the derived performance kernel
and results of simulations of the model in a small system.

\FIGURE{\epsfig{file=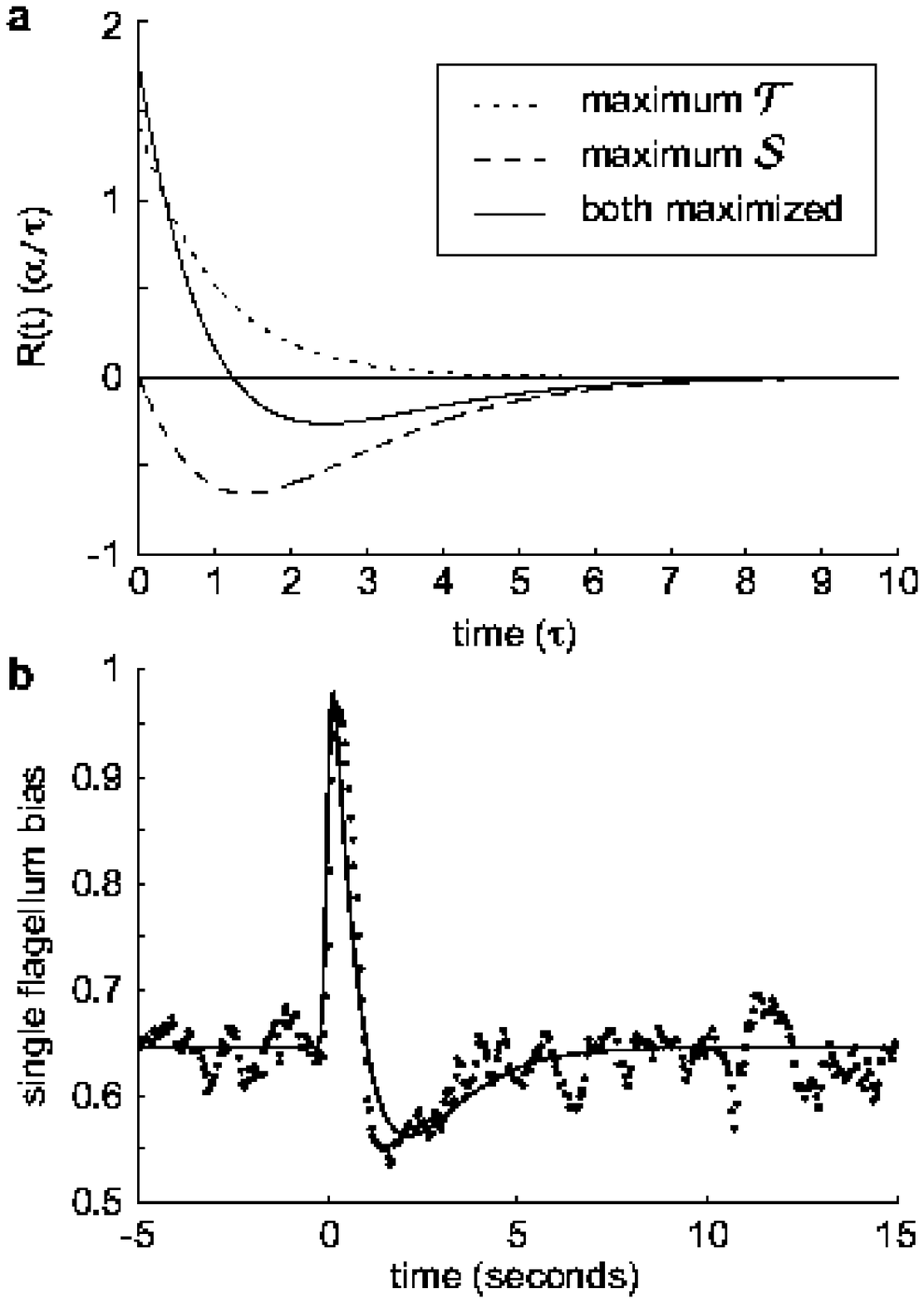,width=9cm}
        \caption{\textsf{\textbf{a}} Response functions that
optimize the performance measures $\mathcal{T}$, $\mathcal{S}$, and
$\mathcal{T}+A\mathcal{S}$, where $A=1/2$. Note that all three
functions are normalized such that $\int R(t)^2 dt = \alpha^2/\tau$.
\textsf{\textbf{b}} The points are data from Figure 1 in
\cite{segall} showing the counter clockwise bias in flagellar motor
rotation after a very short impulse of chemoattractant at time
$t=0$.  The bias response is linear in this experiment's regime. The
solid line is a best fit of $R_{\mathcal{S},\mathcal{T}}(t)$ to the
data, using a 10 Hz low-pass Gaussian filter to realistically smooth
discontinuities. The fitting parameters were $A$, $\tau$, and an
overall amplitude, and the least squares fit was $A=0.56$ and
$\tau=0.9$ seconds.  The bias of a single flagellum is related to
the tumbling probability $P[x(t');t]$, but is not identical because
multiple flagella are involved in running and tumbling \cite{turner}
and cooperative effects could be involved.}%
    \label{figdata}}

%------------------------------------------------------------------
\section{Optimizing the Response Function}
%\textbf{Optimizing the Response Function}

The response functions resulting from optimizing the two performance
criteria have opposite signs, so that optimizing $\mathcal{T}$ leads
to an unfavorable $\mathcal{S}$ and \textit{vice versa}. Both
aspects of performance are biologically relevant -- bacteria should
move up gradients when not in steady state and remain at high
concentrations as they approach steady state. We expect bacteria to
optimize a composite criterion that preserves both aspects of
performance. One can imagine a variety of ways to maximize a
combination of the two quantities, but maximizing any positive
increasing function of both $\mathcal{T}$ and $\mathcal{S}$ will
produce a solution that is a linear combination of $R_\mathcal{S}$
and $R_\mathcal{T}$.  We therefore adopt the most straight forward
way and maximize the quantity
\begin{equation}
\mathcal{T}[R(t)] + A\hspace{1 mm} \mathcal{S}[R(t)]
\end{equation}
where $A$ is some unknown weighting factor of the two performance
measures. As before, we constrain $R(t)$ and take the functional
derivative of this equation to find a response function that
compromises between maximizing the expected run length up gradients
and the steady state bacterial distribution. That response function
is
\begin{equation} \label{eqoptresult}
R_\mathcal{T,S}(t)=\frac{\alpha}{\tau}N_\mathcal{T,S}\exp\{-t/\tau\}\left(1-A(t/\tau+t^2/2\tau^2)\right)
\end{equation}
which is proportional to $R_\mathcal{T}+A R_\mathcal{S}$.

It is reasonable to set $\int R = 0$ because there are physical
bounds placed on the run length of real bacteria.  Purcell
\cite{purcell} pointed out that run duration should be chosen at
least large enough so that, for a given $v$, a bacterium outruns the
diffusion of the chemoattractant $c$ during its run. This lower
bound on run duration is roughly 1 second. Further, in real
situations runs longer than roughly 10\textit{s} are turned 90
degrees off-course by rotational diffusion \cite{brown,block},
setting a maximum useful run duration. Neither of these limits
depends on $c$. Bacteria should be sensitive to gradients by
maintaining a large $\alpha$ but must not allow their run durations
to wander outside these bounds in widely varying concentrations. Run
duration is dependent on a $\nabla c$ term as derived in the text
and on $c \int R(t)dt$. The integral of $R$ should be zero to allow
for sensitivity to $\nabla c$ while keeping $\tilde{\tau}$ within
the limits above, thus creating a large dynamic range for the
response \cite{block}. This argument leads us to set $A = 1/2$ so
that $\int R = 0$. Experimentally, Alon \textit{et al.} \cite{alon}
have shown that this robustness of run duration to changes in
absolute concentration is a property of the \textit{E. coli}
chemotactic network when cells respond to aspartate (though see
\cite{brown}).

The optimized response function is shown in Figure
\ref{figdata}\textsf{\textbf{a}}, where we have required $\int R =
0$.  It predicts a sharp positive immediate response with a drawn
out negative response peaking around $2.5\tau$. It was obtained here
by developing a response function that (1) maximizes the transient
velocity of bacteria up gradients when they are not in steady state
and that (2) creates a steady state where bacteria aggregate in high
concentration regions. The initial, short, positive lobe in
$R_\mathcal{T,S}(t)$ makes $\mathcal{T}>0$ and serves to move the
bacteria up gradients when not in steady state, while the second,
longer, negative lobe makes $\mathcal{S}>0$ and serves to produce
the advantageous steady state distribution.

The functional form of Equation (\ref{eqoptresult}) fits the actual
response function exhibited by individual flagellar motors in Segall
\textit{et al.} \cite{segall} (see Figure
\ref{figdata}\textsf{\textbf{b}}). Our theory concerns the whole
cell, not single flagellar motors; correlations between the activity
of single flagella and the behavior of the whole bacterium are not
well understood \cite{ishihara, turner}.  Nevertheless, we find a
surprisingly good fit.  We have left $A$ and $\tau$ as fitting
parameters, and the best fit yields $A = 0.56$, which matches our
expectation that $A \simeq 0.5$.

Values of $\mathcal{T}$ and $\mathcal{S}$ for any response function
can be easily found by calculating their overlap with the kernels in
Equations (\ref{eq_T_definition}) and (\ref{S_definition}); a
summary of such calculations for our three optimizations is shown in
Table 1. The top half of the table provides the qualitative picture
independent of model details, while the lower half provides the
values of $\mathcal{T}$ and $\mathcal{S}$ given by our model.

\begin{table}[h]
\caption{\label{tabequations} Transient velocity values
($\mathcal{T}$) and steady state strength of aggregation
($\mathcal{S}$) for the various response functions $R(t)$. The first
set is heuristic, derived from qualitative arguments, while the
second set is derived from our particular model. The first item in
each section is the response function maximizing $\mathcal{T}$,
while the second maximizes $\mathcal{S}$. The third maximizes both
$\mathcal{T}$ and $\mathcal{S}$, as described in the text. More
positive values of $\mathcal{T}$ and $\mathcal{S}$ indicate more
favorable behavioral performance; in the heuristic section,
favorable values are represented by ($+$) and unfavorable by ($-$).
} \center
\begin{tabular}{r|l|r@{.}l r@{.}l}
Response Function & Equation & \multicolumn{2}{c}{$\mathcal{T}$} &
\multicolumn{2}{c}{$\mathcal{S}$} \\
 \hline\hline $R_\mathcal{T}(t)$ & Positive lobe, weighted
towards $t=0$ & \multicolumn{2}{c}{$+$} &
\multicolumn{2}{c}{$-$} \\
$R_\mathcal{S}(t)$ & Negative lobe, weighted towards $t\simeq\tau$ & \multicolumn{2}{c}{$-$} & \multicolumn{2}{c}{$+$} \\
$R_\mathcal{T,S}(t)$ & Initial brief positive lobe; negative lobe peaked beyond $\tau$ & \multicolumn{2}{c}{$+$} & \multicolumn{2}{c}{$+$} \\
 \hline $R_\mathcal{T}(t)$ & \hspace{2.5mm}$\frac{\alpha}{\tau}N_\mathcal{T}e^{-t/\tau} $  &
$0$ & $7$ & $-0$ & $5$ \\
 $R_\mathcal{S}(t)$ & $ -\frac{\alpha}{\tau}N_\mathcal{S}\left(t/\tau+t^2/2\tau^2\right)e^{-t/\tau}$ & $-0$ & $4$ & $0$ & $9$ \\
 $R_\mathcal{T,S}(t)$ & \hspace{2.5mm}$\frac{\alpha}{\tau}N_\mathcal{T,S}\left(1-\frac{1}{2}(t/\tau+t^2/2\tau^2)\right)e^{-t/\tau}$  & $0$
& $5$ & $ 0$ & $05$ %\\
\end{tabular}
\end{table}

%------------------------------------------------------------------------------
\section{Discussion}
%\textbf{Discussion}

The biphasic shape of the chemotactic response function has long
been interpreted as a temporal comparator that automatically adapts
to offsets in chemical concentration \cite{segall}. Here we have
examined two aspects of bacterial chemotactic performance and found
that neither aspect optimized alone produces a biphasic response. A
composite response function, simultaneously optimizing both measures
of performance, closely fits the shape of the experimental data.
This leads to an additional interpretation of $R(t)$ as optimized
with respect to these two behaviors, connecting each lobe to
distinct behavioral performance.

Our theory makes novel predictions about the behavior of wild-type
and mutant bacteria. The functional fit of $R_\mathcal{T,S}$ to the
wild-type data is quite good, so that we predict that experimental
measures of wild-type $\mathcal{T}$ and $\mathcal{S}$ in the linear
regime should roughly match those in Table 1. If $L$ is the decay
length of $b(x)$ when bacteria are on a linear gradient, it is
related to the expected transient velocity up the gradient, $v_d$,
by the relation $L=(v^2\tau/v_d) *(\mathcal{T}/\mathcal{S})$. The
first factor could be found on dimensional grounds, but we predict
$\mathcal{T}/\mathcal{S} \simeq 14$ for wild-type \textit{E. coli}.
For a gradient that elicits $v_d=1\mu/$\textit{s}, this predicts
$L\simeq5$ millimeters. Response functions of mutant bacteria can be
calculated (see \cite{tu} or \cite{shimizu}) or measured
experimentally, as Segall \textit{et al.} \cite{segall} have done
for strains with mutant \textit{cheZ} and for strains with mutant
\textit{cheRcheB}. Both these mutant response functions are entirely
positive with durations of roughly 5 and 1 seconds, respectively. We
predict that both mutants will have transient velocities up
gradients, but that both will reach an unfavorable steady state
distribution. Available data for both mutants does not rule out
these predictions \cite{boesch, parkinson, stock1985}. Microscopic
observations of $\overline{t}^+ - \overline{t}^-$ or measurements of
$b(x)$ in static spatial gradients could evaluate the validity and
limits of this theory.

\section{Acknowledgements}  We benefitted from fruitful conversations
with and guidance from Howard Berg, Yariv Kafri, Aravinthan Samuel,
Tom Shimizu, and Rava da Silveira. D.A.C. is funded by the NSF and
L.C.G. is funded by the DOE.

\appendix

\section{Steady State and Tumbling Probabilities}
Here we derive
Equation (\ref{eqssfundamental}), the steady state distribution of
bacteria in terms of the expected tumbling probabilities of bacteria
coming from the right and left.

In the steady state, the net flux of bacteria through a point
between $x$ and $dx$ must be zero.  Bacterial concentration at point
$x$ is represented by $b^\pm(x)$, where the ($\pm$) indicates the
direction of movement. Setting the net flux to zero yields
\begin{eqnarray}
\label{eqflux}
b^+(x-dx)(1-\frac{1}{2}  \overline{P}^+(x) \frac{dx}{v})+b^-(x-dx)\frac{1}{2} \overline{P}^-(x) \frac{dx}{v}=\\
b^-(x+dx)(1-\frac{1}{2} \overline{P}^-(x)
\frac{dx}{v})+b^+(x+dx)\frac{1}{2} \overline{P}^+(x) \frac{dx}{v}
\nonumber
\end{eqnarray}
The left hand side counts bacteria moving up the gradient past $x$,
while the right hand side counts bacteria moving down the gradient
past $x$. The probability of actually reversing directions is half
the probability of tumbling. The first term on each side represents
bacteria continuing on their present course; the second represents
bacteria passing through the point after an instantaneous
reorientation. Retaining only zeroth order terms gives
$b^+(x)=b^-(x)$, so one can replace each by $b(x)/2$. Expanding
$b(x)$ about $x$ and retaining only terms up to first order in $dx$
yields a differential equation governing the steady state
distribution of bacteria:
\begin{equation} \label{eqssfundamental_diff_app}
\frac{\nabla b(x)}{b(x)}={\overline{P}^-(x) - \overline{P}^+(x)\over
2v}
\end{equation}
This is integrated to find Equation (\ref{eqssfundamental}). The
zero flux condition holds at all points in the system.  The equation
is modified at system boundaries, depending on conditions there, but
the first order differential Equation
(\ref{eqssfundamental_diff_app}) describes the bacterial
distribution far from the boundaries.

%-----------------------------------------------------------------------
\section{Optimizing the Steady State Distribution}

Here we calculate $\overline{P}^-(x) - \overline{P}^+(x)$ in the
steady state and find a response function that optimizes the
bacterial distribution, $b(x)$.

We are interested in:
\begin{equation} \label{eqPtdiffdef}
\overline{P}^-(x) - \overline{P}^+(x)
=\frac{1}{\tau}\int_{-\infty}^0 dt'
R(-t')\big[\overline{c}^+(t')-\overline{c}^-(t')\big]
\end{equation}
where superscripts indicate that the averages are taken over all
paths ending at the position $x$ moving either up ($+$) or down
($-$) the gradient at $x$.  We take the gradient to be positive, so
that up-moving bacteria are right-moving bacteria.

To calculate this quantity, we must insert averages over possible
paths. Assuming that the most recent tumble occurred at time $t_0$,
the one before that at time $t_1$, and so on, we will average over
the values of $t_0,t_1,\ldots$ and over the directions of the runs
during each of the intervals.  Essentially, because
$\overline{c}^\pm$ is multiplied by $R(t)$ in the integral, this
calculation can neglect first order effects of $R(t)$ on
$\overline{c}^\pm$: runs become simply exponential in length and
there is an equal probability that a bacterium came from left or
right before its last tumble.

We first average over the directions of the runs. In the steady
state, for a tumble at $t_i$, the probability $Q^+(t_i)$ that the
bacterium was moving up the gradient before the tumble will be
proportional to the population of bacteria found to the left of
$x(t_i)$. That is
\begin{equation}
Q^+(t_i) = \frac{b(x(t_i)-v(t_i-t_{i+1}))}{b(x(t_i)-v(t_i-t_{i+1}))+
b(x(t_i)+v(t_i-t_{i+1}))}
\end{equation}

Now assume that $b(x)$ varies slowly over one run and expand $b(x)$
to first order in $t_i-t_{i+1}$.  Then
\begin{equation}\label{Qout}
Q^+(t_i) \simeq \frac{b(x(t_i))-\nabla b v (t_i-t_{i+1})}{2
b(x(t_i))}=\frac{1}{2}\bigg(1-v\frac{\nabla b}{b}(t_i-t_{i+1})\bigg)
\end{equation}

Inserting  just the averages over the directions of motion in the
various intervals and leaving the $t_i$ fixed for the moment, we can
expand Equation (\ref{eqPtdiffdef}) as:
\begin{eqnarray}\label{vaverage}
& \int_{-\infty}^{0} dt'R(-t')\overline{c}^{\pm}(t') \to  \\
& \int_{t_0}^{0} dt'R(-t')c^{\pm}(t') \nonumber \\
& + \int_{t_1}^{t_0} dt'R(-t')\left[Q^+(t_0)
c^{(\pm+)}(t')+Q^-(t_0)c^{(\pm-)}(t')\right] \nonumber
\\ & + \int_{t_2}^{t_1} dt'R(-t')\big[Q^+(t_0)Q^+(t_1) c^{(\pm++)}(t')+Q^+(t_0)Q^-(t_1)
c^{(\pm+-)}(t') \nonumber \\
&+Q^-(t_0)Q^+(t_1) c^{(\pm-+)}(t')+Q^-(t_0)Q^-(t_1)
c^{(\pm--)}(t')\big] + \ldots \nonumber
\end{eqnarray}
Here $c^{++-}(t)$, for instance, denotes the concentration seen by a
bacterium that has moved in the $+$ direction after $t_0$, moved in
the $+$ direction in the interval $[t_1,t_0]$, and in the $-$
direction in the interval $[t_2,t_1]$. From Equation
(\ref{eqssfundamental_diff_app}), $\nabla b/b=(\overline{P}^- -
\overline{P}^+)/v$, which is a sum of terms proportional to
integrals of the form $\int R(t-t')c(t')dt'$, so we may set
$Q^{\pm}(t_i) = 1/2$ to keep only terms to first order in $\int
R(t-t')c(t')dt'$.

Now we look at the averages in square brackets in Equation
(\ref{vaverage}). In the first such average, we expand $c(t)$ about
$t_0$ to find that
\begin{eqnarray} \label{Qave}
& Q^+(t_0)c^{(\pm+)}(t')+Q^-(t_0)c^{(\pm-)}(t')  \\ &
=1/2(c^{\pm}(t_0)+v (t'-t_0)\nabla c+c^{\pm}(t_0)-v (t'-t_0)\nabla
c)\nonumber
\\ & = c^{\pm}(t_0) \nonumber
\end{eqnarray}
The second square bracket in (\ref{vaverage})reduces to
$(c^{(\pm+)}(t_1) + c^{(\pm-)}(t_1))/2$ and the rest of the square
brackets will reduce similarly.

Inserting the averages over the tumbling times $t_0, t_1, \ldots$,
Equation (\ref{eqPtdiffdef}) expands to:
\begin{eqnarray}\label{taverage}
&\int_{-\infty}^{0} dt'R(-t')\overline{c}^{\pm}(t') \to \\
&\int_{-\infty}^0 dt_0 D(t_0|0) \int_{t_0}^{0}
dt'R(-t')c^{\pm}(t') \nonumber \\
& + \int_{-\infty}^{0}dt_0 D(t_0|0)\int_{-\infty}^{t_0}
dt_1 D(t_1|t_0)\int_{t_1}^{t_0} dt'R(-t')c^{\pm}(t_0) \nonumber \\
& + \int_{-\infty}^{0}dt_0 D(t_0|0)\int_{-\infty}^{t_0} dt_1
D(t_1|t_0) \int_{-\infty}^{t_1} dt_2 D(t_2|t_1) \nonumber \\
& \times \int_{t_2}^{t_1}
dt'R(-t')\frac{(c^{(\pm+)}(t_1)+c^{(\pm-)}(t_1))}{2} + \ldots
\nonumber
\end{eqnarray}
where $D(\theta_2 | \theta_1)$ is the probability that the bacterium
tumbled at $\theta_2$ given that it tumbled later at $\theta_1$,
where $\theta_2 < \theta_1$ so that we are reconstructing the
tumbles backwards in time.

It remains to write out the factors $D(t_{i+1}|t_i)$ explicitly in
terms of $R(t)$.  Given the model,
\begin{equation}
D(t_{i+1}|t_i)=\frac{\exp\bigg\{-\int\limits_{t_{i+1}}^{t_i} dt'
P[x(t'');t']\bigg\}}{\int\limits_{-\infty}^{t_i} dt_{i+1}
\exp\bigg\{-\int\limits_{t_{i+1}}^{t_i} dt' P[x(t'');t']\bigg\}}
\end{equation}
where the expression is normalized to integrate to $1$. Keeping only
terms up to first order in $R(t)$ allows us to set
$D(t_{i+1}|t_i)=1/\tau \exp\{(t_{i+1}- t_i)/\tau\}$.  Then
$\overline{P}^-(x) - \overline{P}^+(x)$ becomes
\begin{eqnarray} \label{cutoff}
& \overline{P}^-(x) - \overline{P}^+(x)=
\\
& \int_{-\infty}^0 {dt_0\over\tau} e^{t_0/\tau} \int_{t_0}^{0}
{dt'\over\tau} R(-t')\big[c^+(t')-c^-(t')\big] \nonumber \\
& + \int_{-\infty}^0 {dt_0\over\tau} \int_{-\infty}^{t_0}
{dt_1\over\tau} e^{t_1/\tau} \int_{t_1}^{t_0} {dt'\over\tau}R(-t')\big[c^+(t_0)-c^-(t_0)\big] \nonumber \\
& + \int_{-\infty}^0 {dt_0\over\tau} \int_{-\infty}^{t_0}
{dt_1\over\tau}
\int_{-\infty}^{t_1} {dt_2\over\tau} e^{t_2/\tau} \nonumber \\
& \times \int_{t_2}^{t_1} {dt'\over\tau}
R(-t')\frac{\big[c^{(++)}(t_1)+c^{(+-)}(t_1)-c^{(-+)}(t_1)-c^{(--)}(t_1)\big]}{2}
+ \ldots \nonumber
\end{eqnarray}

First consider the quantities in square brackets.  The first one can
be approximated by $c^+(t')-c^-(t')=2v t' \nabla c $. The second one
can also be approximated the same way, but $t_0$ replaces $t'$.
Whether the third one can be made proportional to the gradient of
the concentration depends on how quickly the gradient varies in
space. Several of the terms in (\ref{cutoff}) might be approximated
in terms of $\nabla c$; it is a question of what sorts of gradients
a bacterium typically encounters.

Consider a generic chemical landscape that is approximately flat on
a large length scale $L$. On length scales smaller than $L$ the
concentration may vary significantly.  In such a landscape,
quantities like those in square brackets in (\ref{cutoff}),
representing measurements made in the distant past ($|t'| \gg
{L^2\over v^2\tau}$) will have a tendency to sum to zero.  In
particular, if $L \sim 2 v\tau$, the quantity in the third square
bracket in (\ref{cutoff}) will be close to zero since on average

\begin{equation}
c^{(++)}(t_1)+c^{(+-)}(t_1) \sim c^{(-+)}(t_1)+c^{(--)}(t_1)
\end{equation}

Terms coming from the past where $|t'| \gg {L^2 \over v^2\tau}$
cease to contribute to $\overline{P}^-(x) - \overline{P}^+(x)$.
Where we cut off the series in Equation (\ref{cutoff}) is a
biological question. We expect bacterial strategy to make minimal
assumptions about the extent of the gradient, and therefore we will
cut off the series after only a few terms. We drop all terms that
refer to times earlier than $t_1$. The resulting expression for
$\overline{P}^-(x) - \overline{P}^+(x)$ has two terms proportional
to $\nabla c$. The terms proportional to $\nabla c$ are precisely
the ones that allow us to optimize the steady state distribution in
a concentration independent way.

We find that
\begin{eqnarray} \label{cutoffresult}
& \overline{P}^-(x) - \overline{P}^+(x) =\frac{2v \nabla c}{\tau^2}
\big[\int_{-\infty}^0 dt_0 e^{t_0/\tau} \int_{t_0}^0 dt' R(-t')t'
 \\  & + \int_{-\infty}^0 dt_0 \int_{-\infty}^{t_0} dt_1
e^{t_1/\tau} \int_{t_1}^{t_0} \frac{dt'}{\tau} R(-t') t_0 \nonumber
\big]
\end{eqnarray}
Letting $R(t)= \int_0^{\infty} ds R(s) \delta (t-s)$, we find the
simple result that $\overline{P}^-(x) - \overline{P}^+(x)$ can be
written as the overlap in Equation (\ref{eqPtdiff}).

Keeping more terms in Equation (\ref{cutoff}) corresponds to
assuming that the bacteria propagate on gradients with variations of
longer length scale. Such an assumption produces a kernel of
$e^{-t/\tau}$ multiplied by an expansion of $(1-e^{t/\tau})$ in
$t/\tau$. Retaining an infinite number of terms leads to a kernel
proportional to $e^{-t/\tau}-1$. As long as a finite number of terms
are retained, the qualitative features of the performance kernel
don't change: it starts at zero, peaks negatively, and returns to
zero for large $t$. If the number of retained terms remains small
(less than 5, for example), the quantitative features are not much
affected either.

Figure \ref{appfig} illustrates this discussion. It shows the
results of simulations of the model that demonstrate the
contribution to $\mathcal{S}$ from each portion of the response
function. The response functions are chosen as in Figure
\ref{figsim} to weight only $c(t-\theta)$ in determining the turning
probability. Bacteria navigating concentration gradients of various
length scales are considered. On the timescales shown, the white
points in the figure represent bacteria in an effectively infinite
linear gradient. Bacteria heading down such a gradient at $x$ on
average have a higher $c$ history for \textit{all previous times}
than ones heading up the gradient at $x$. Therefore, measurements
made any time in the past, including long ago, affect $\mathcal{S}$
by the same amount. On the other hand, when the gradient is not
infinite, as in any real case, $\nabla c$ will change on some length
scale. Measurements of $c(t)$ made with a delay large compared to
the time to traverse this length scale will average out in the sum
over histories. Such measurements therefore cease to affect
$\mathcal{S}$. As the grey points in the figure show, on a
relatively short length scale, measurements of concentration made
$20 \tilde{\tau}$ in the past no longer contribute to $\mathcal{S}$.
In the time of $20 \tilde{\tau}$, the bacterium is likely to have
bounced against a wall (or moved over a peak, if one views the
reflecting box as an infinite triangle wave), so such measurements
no longer reflect the bacterium's current gradient. With still
shorter gradients, such effects become more pronounced. The black
points in the figure are on a length scale that roughly matches our
theory, in which we restricted bacteria to looking at only the
previous two runs. The simulation conditions mean that bacteria
cannot use measurements made long ago, while our theory posits that
bacteria should make minimal assumptions about gradient length. Both
lead to similar kernels showing the influence of $R(\theta)$ on
$\mathcal{S}$.

\FIGURE{\epsfig{file=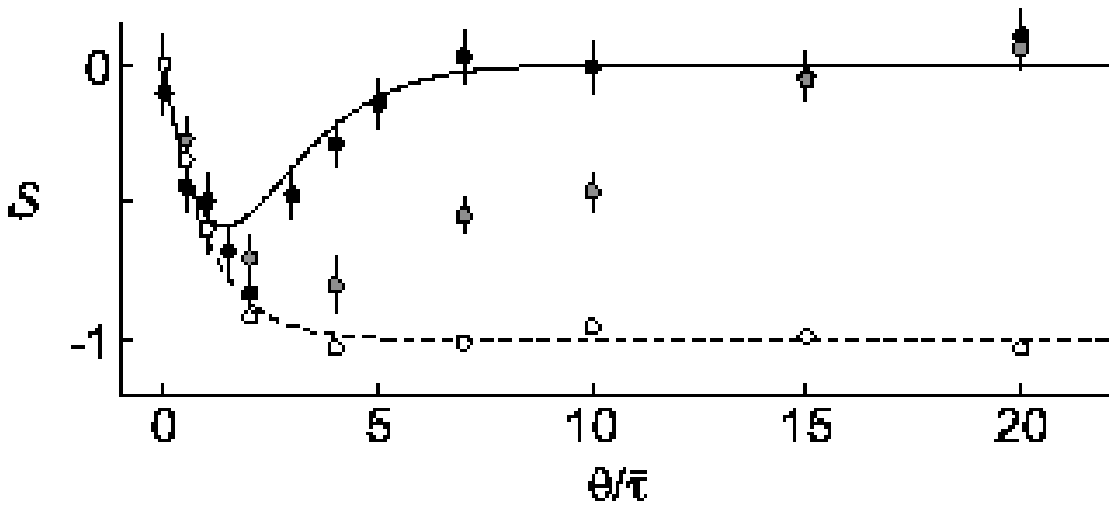,width=10cm}
        \caption{Using the same parameters as in Figure
\ref{figsim}\textsf{\textbf{d}}, we here show the results on
different length scale gradients.  Such gradients with reflective
boundary conditions can also be thought of as infinite triangle
waves, a more natural picture than a box. Black points are
simulations of bacteria in a box of total length $4 v\tilde{\tau}$,
grey points are in a box of length $8 v\tilde{\tau}$, and white
points are in a box of length $80 v\tilde{\tau}$.  The dotted curve
shows the performance kernel for $\mathcal{S}$ in the case where the
bacterium assumes a linear gradient throughout its entire history
(an infinite gradient). The solid curve is the performance kernel
that makes minimal assumptions about the extent of the linear
gradient.}%
    \label{appfig}}

\end{document}